# Blue (In,Ga)N Light-Emitting Diodes with Buried $n^+$-$p^+$ Tunnel Junctions by Plasma-Assisted Molecular Beam Epitaxy

YongJin Cho,[1,a)] Shyam Bharadwaj,[1] Zongyang Hu,[1] Kazuki Nomoto,[1] Uwe Jahn,[2] Huili Grace Xing,[1,3] and Debdeep Jena[1,3,b)]

[1]School of Electrical and Computer Engineering, Cornell University, Ithaca, New York 14853, USA
[2]Paul-Drude-Institut für Festkörkperelektronik, Hausvogteiplatz 5–7, 10117 Berlin, Germany
[3]Department of Materials Science and Engineering and Kavli Institute for Nanoscale Science, Cornell University, Ithaca, New York 14853, USA

ABSTRACT

Blue light-emitting diodes (LEDs) consisting of a buried $n^+$-$p^+$ GaN tunnel junction, (In,Ga)N multiple quantum wells (MQWs) and a $n^+$-GaN top layer are grown on single-crystal Ga-polar $n^+$-GaN bulk wafers by plasma-assisted molecular beam epitaxy. The (In,Ga)N MQW active regions overgrown on the $p^+$-GaN show chemically abrupt and sharp interfaces in a wide range of compositions and are seen to have high structural and optical properties as verified by X-ray diffraction and spatially resolved cathodoluminescence measurements. The processed LEDs reveal clear rectifying behavior with a low contact and buried tunnel junction resistivity. By virtue of the top $n^+$-GaN layer with a low resistance, excellent current spreading in the LEDs is observed at low currents in this device structure. A few of new device possibilities based on this unique design are discussed.

---

a) Electronic mail: yongjin.cho@cornell.edu
b) Electronic mail: djena@cornell.edu

Compared to narrower bandgap compound semiconductors such as the group III-Arsenides, the group III-Nitrides possess relatively large ionic bonding nature stemming from large differences in electronegativity between the cations and N, which results in the superior optical properties of the material system. Combined with wide span of direct bandgaps, the group III-Nitrides are thus attractive for and have been actively utilized in light emitting devices.[1] For technical reasons, most nitride light emitting diodes (LEDs) so far have focused on heterostructures with *p*-GaN layer as the top layer, grown on top of the optically active layers. Especially metal organic chemical vapor deposition (MOCVD)-grown nitride LEDs need a post-growth-annealing step in order to break Mg-H bonds and activate *p*-GaN:Mg layers, requiring the *p*-GaN layer to be located on the surface. Due to the relatively high resistivity of *p*-GaN, however, current spreading is problematic in such structures. Contrarily, MBE does not require a post-growth-annealing-step for activating *p*-GaN:Mg since the growth is performed under H-free-ultra-high-vacuum environment, and therefore there is no penalty for positioning buried *p*-GaN:Mg layer in MBE structures.

On the other hand, wurtzite III-Nitride semiconductor heterostructures exhibit strong spontaneous and piezoelectric polarization fields of the order of a few MV/cm along the polar *c*-axis.[2] These polarization fields cause quantum-confined Stark effect in the active regions of quantum-well LEDs. The polarization-induced reduction of the oscillator strength due to poor electron-hole overlap reduces the efficiency of LEDs. However, polarization engineering in such heterostructures offers several creative opportunities for photonic and electronic devices such as tunnel junctions, including ultra-low power tunneling transistors.[3,4,5] The N-polar direction has recently drawn attention for unique device properties such as buried-barrier HEMTs and interband tunnel junctions.[6,7,8,9,10] Although epitaxial growth along the N-polar direction presents certain fundamental advantages stemming from the polarity-dependent decomposition temperatures of the materials,[11] it has been reported that N-polar nitrides show considerably low luminous efficiency and are also vulnerable to wet chemical etching process, preventing the realization of the full advantage of the favorable polarization fields in N-polar nitrides.[12,13]

In this paper, Ga-polar (In,Ga)N LED structures with a buried $p^+$-GaN and a top $n^+$-GaN were grown on GaN(0001) bulk wafers. This is structurally equivalent to the conventional (In,Ga)N LED structures, i.e., (substrate) / $n$-GaN / (In,Ga)N / $p$-GaN (surface), grown along the N-polar direction. Thus, the structure in this study enables to not only keep the high luminescence properties of Ga-polar (In,Ga)N but also exactly *mimic* the polarization properties of N-polar nitrides, while showing excellent current spreading on the top surface due to the $n^+$-GaN cap layer. This novel idea was introduced in a recent work with single (In,Ga)N emitter regions, and fixed doping densities.[5] In this work, we introduce multiple (In,Ga)N quantum wells, and explore the dependence of the behavior of the buried tunnel junctions with doping. Furthermore, we explore by spatially resolved cathodoluminescence the optical emission in the active regions, and demonstrate uniform light emission over mm-scale diodes at low current densities using the new device geometry.

The (In,Ga)N LED structures were grown on single-crystal Ammono Ga-face GaN(0001) bulk wafers with a dislocation density of $5\times10^4$ cm$^{-2}$ in a Veeco Gen10 MBE reactor equipped with standard effusion cells for elemental Ga, In, Mg and Si, and a radio-frequency plasma source for the active N species. Si was used as the $n$-type donor, and Mg as the $p$-type acceptor for doping of the GaN layers. The base pressure of the growth chamber was in the range of 10$^{-10}$ Torr under idle conditions, and $2\times10^{-5}$ Torr during the growth runs. The MBE-grown (In,Ga)N LED structures starting from the nucleation surface is 200 nm GaN:Si / 100 nm GaN:Mg / 5 period (In,Ga)N multiple quantum wells (MQWs) / 100 nm GaN:Si. The details of the layer structures are shown in Fig. 1(a). The $p^+$-GaN layer is located below the active region and forms a tunnel junction (TJ) with the heavily doped $n^+$-GaN layer underneath it. Two LED samples differing in Mg concentrations [$5\times10^{18}$ cm$^{-3}$ (sample A) and $3\times10^{19}$ cm$^{-3}$ (sample B)] of the $p^+$-GaN layers were prepared in order to study the impact of the Mg concentration on the performance of the LED. All the GaN layers ($\Phi_{Ga} > \Phi_N$) and the (In,Ga)N MQWs ($\Phi_{In} + \Phi_{Ga} > \Phi_N$; $\Phi_{Ga} < \Phi_N$) were grown under metal-rich conditions at 710 ℃ and 660 ℃, where $\Phi_{Ga}$, $\Phi_{In}$ and $\Phi_N$ are Ga, In and active N fluxes, respectively. The growth rate, which is limited by $\Phi_N$, was 7 nm/min. Two Ga effusion cells were used for the growth of the (In,Ga)N MQWs in order to obtain abrupt changes in the Ga fluxes in the well ($\Phi_{Ga}$=5.5 nm/min) and the barrier layers ($\Phi_{Ga}$=6 nm/min), which is known to be critical for the MBE growth of high-quality (In,Ga)N QWs.[14] The excess Ga droplets after the growth were first removed in HCl before ex situ characterization and device fabrication. The (In,Ga)N LED samples were fabricated by optical contact-lithography followed by two-step mesa etching in Cl$_2$- and BCl$_3$-based inductively-coupled plasma (ICP) etching. For alloyed ohmic contacts on the top and bottom n$^+$-GaN surface, the same Ti (25 nm) / Al (100 nm) metal stack was deposited in N$_2$ ambient at 550 ℃ for 1 minute by DC-sputtering. Structural properties and surface morphology of the samples were characterized by in situ reflection high energy electron diffraction (RHEED), atomic force microscopy (AFM) and x-ray diffraction (XRD) measurements. Spatially resolved cathodoluminescence (CL) spectroscopy using a beam energy of 7 keV in a scanning electron microscope (SEM) was used to probe the local variations of the optical properties of the samples.

Electrical transport and electroluminescence (EL) measurements were used to evaluate the device performance. All the ex situ characterization steps were performed at room temperature.

Figure 1(b) displays a RHEED pattern of sample A taken at low temperature (< 300℃) after growth. The formation of a laterally contracted metallic Ga bilayer structure is confirmed by the observation of the diffused satellite streaks ("1×1" reconstruction) in the RHEED pattern as indicated by the two arrows in Fig. 1(b). These streaks guarantee that there was no polarity inversion during the growth.[15,16] The AFM micrograph shown in Fig. 1(c) reveals the smooth surface morphology exhibiting clear atomic steps. Note that there are no spiral hillocks on the surface, which are commonly observed on group III-Nitride layers grown by MBE on substrates with high dislocation densities,[17] indicating that dislocation density in this sample is low and no strain relaxation occurred during the growth of the (In,Ga)N active region. Instead the surface morphology is characterized by wide step terraces with widths of 300 – 500 nm. Figure 1(d) shows the symmetric XRD ω-2θ scan of sample A. Excellent agreement between the simulated curve based on the layer structure in Fig. 1(a) and the experimental data can be seen, implying that the interfaces are sharp in a wider range and each layer in the structure is chemically abrupt.

The indium distribution and luminescence characteristics of the (In,Ga)N in a micro-scale was studied by using spatially resolved cathodoluminescence (CL). Figure 2(a) shows the spatially averaged CL spectrum of sample A. Two clear CL peaks, one from GaN at ~365 nm and the other from the (In,Ga)N MQWs at 444 nm are clearly seen. The single CL peak with a narrow full-width-at-half-maximum (FWHM) of 120 meV from the (In,Ga)N MQWs implies that there is no phase separation in the MQWs and each (In,Ga)N well and barrier layer is compositionally homogeneous.

This can also be inferred from the excellent match between the XRD data and the simulation in Fig. 1(d). Figures 2(b)–(f) show secondary electron (SE) [Fig. 2(b)] and CL images [Figs. 2(c)–(f)] for different values of the CL detection wavelength on the same region of sample A. The SE image [Fig. 2(b)] is linked to the morphological features of the AFM image, [Fig. 1(c)] i.e., the surface is characterized by large step terraces and edge structure. It should be noted that the CL images do not show any clear dark spots which are easily found for (In,Ga)N grown on substrates with a high dislocation density.[18] Such dark spots in CL can readily be connected to threading dislocations (TDs) with screw component (*c*-type dislocation: Burgers vector of *c*) which typically induce spiral hillocks on the surface and V-pits for (In,Ga)N.[18] Such TDs with screw component are especially more detrimental than other defects for (In,Ga)N-based optical devices as they not only act as strong non-radiative centers but also induce lateral fluctuations of In incorporation around them resulting in broadening the optical transition energy.[18]

Figures 2(d)–(f) show the CL images for different values of detection wavelength corresponding to the high-energy [Fig. 2(d)], the center [Fig. 2(e)], and the low-energy [Fig. 2(f)] side of the (In,Ga)N MQW CL spectrum of Fig. 2(a). Here the spectral resolution of the maps amounted to 5 nm. Within this resolution, overall the CL intensity distribution does not vary clearly with the detection wavelength but with the morphology, not showing any TD-related features (dark spots). This is indeed supported by the fact that not a single spiral hillock on the surface over an area of 20 × 20 µm$^2$ could be observed by AFM (not shown here). Instead, the CL characteristics of this sample is seen to be correlated with the surface structure, i.e., the CL tends to be brighter on the terraces compared to that on the edges regardless of the CL detection wavelength.

In order to study this correlation between the CL and the morphology in more detail, a CL line scan with a higher spectral resolution of 0.5 nm was performed along a line passing through a few atomic step terraces [Fig. 2(g)]. Figure 2(h) shows the corresponding CL map as functions of scan position and wavelength taken along the scan line in Fig. 2(g). Two CL bands from the GaN and the (In,Ga)N MQWs are clearly seen. As can be expected from Figs. 2(c)–(f), the intensities of these CL bands tend to be modulated in phase along the scan position, i.e., the CL intensities of both the GaN and the MQWs are seen to be high on the terraces and low at the edges. Figure 2(i) shows the central CL wavelength and the maximum CL intensity of the (In,Ga)N MQWs as a function of the scan position. One can see that both the CL wavelength and the maximum CL intensities of the (In,Ga)N MQWs tend to be higher on the terraces than at the edges, e.g., see the three terrace regions divided by the three dashed lines in Figs. 2(g)–(i). At the edges, the wavelength tends to be a bit (~1 nm) lower than on the terraces. Thus carriers excited at the edges will drift towards the terraces before radiative recombination. Since for CL only the excitation is spatially resolved and not the detection, high-energy regions (the edges) with neighboring low-energy regions (terraces) show a smaller intensity in CL. The same behavior holds for the GaN [Fig. 2(h)].

Now we turn to the effects of Mg doping density on the LED performance by studying samples A and B which contain different Mg concentrations in the buried $p^+$-GaN layers. We first check the top contacts by using transmission line model (TLM) analysis [Fig. 3(a)]. As expected, the top $n^+$-GaN layers show linear current vs voltage relations with clear scaling effects of the TLM pad spacings [e.g., see the inset in Fig. 3(a) for sample A]. The extracted contact resistivity and sheet resistance are $1.35\times10^{-5}$ $\Omega cm^2$ ($9.13\times10^{-6}$ $\Omega cm^2$) and $1.69\times10^2$ $\Omega$/sq ($1.18\times10^2$ $\Omega$/sq) for sample A (B), respectively. Thus, the two samples show similar low contact resistivities. Figure 3(b) and the inset depict the logarithmic and linear current density vs voltage (J-V) relation of the LEDs, respectively, exhibiting clear rectification for both samples. Here the positive voltage was applied on the bottom contact and the top contact was grounded. It is seen that sample B, which has a higher Mg doping density than sample A, reveals a lower diode turn-on voltage than sample A, indicating that carrier tunneling indeed depends on the doping densities at the $n^+$-$p^+$ TJ.

The apparent turn-on voltages of ~5 V and ~12 V in the linear J-V curves indicate the additive voltage drop incurred at the buried tunnel junction. This turn-on voltage can be lowered further by heavier doping (without compromising the active layer growth), and/or by inserting a thin (Al,Ga)N layer between the $n^+$-$p^+$ junction for polarization-boost for interband tunneling, and/or by decreasing the number of MQWs.[19] The significantly higher turn-on voltage for Sample A is due to a much higher reverse-bias voltage that must drop across the wider depletion region of the buried tunnel junction (BTJ) of the lighter doped Sample A, to drive the same current as the heavier doped Sample B. In other words, while ~2 V drops across the BTJ of sample B, ~8 V is needed in sample A. This indicates a strong dependence of the voltage drop on the Mg doping of the BTJ. In addition, the turn-on at TJs may become more sensitive to dopant distribution with lower doping density, which may lead to less current spreading on the device. The total series resistance of the device in sample B in forward bias has been extracted by fitting the linear region of the forward bias J-V curve to be $8.35\times10^{-3}$ $\Omega cm^2$, indicating that the contact resistance ($9.13\times10^{-6}$ $\Omega cm^2$) of the top $n$-type contact is negligible in this total resistance. The specific resistivity of the $n^+$-$p^+$ junction is therefore lower than $8.35\times10^{-3}$ $\Omega cm^2$, which can be comparable with the lowest resistances observed for nitride-based tunnel junctions, the difference being the samples reported here have little or no dislocations that can provide additional leakage paths.[20,21,22]

Figure 3(c) displays the normalized electroluminescence (EL) intensity spectra of the two devices in Fig. 3(b). The EL spectra consist of a single peak located around 444 nm. It is seen that sample B shows a slightly broader EL peak compared to sample A, implying that higher Mg incorporation induces a slight crystal degradation of the overgrown diode. It can also be judged from the higher leakage current of sample B than sample A in reverse bias [Fig. 3(b)]. The insets in Fig. 3(c) show optical micrographs taken from a large device with a size of 0.5×0.1 mm$^2$ of sample B at different injection currents. As is expected for the buried-tunnel junction geometry, one can see that current spreading in the $n^+$-GaN layer is excellent, enabling light emission from the whole device area even at a low injection current density. For the higher resistance sample A, the current spreading is not as efficient (not shown here) as sample B, underlining the need for higher Mg doping in the buried TJs.

To conclude, the buried-tunnel junction (In,Ga)N LED structures reported in this study not only show excellent current spreading, but also has several other advantages due to the inherent properties where polarization and *p-n* junction fields are aligned parallel to each other. Wavefunctions of carriers are more concentrated in the well regions and electron injection and blocking in the active region are more facilitated, making emission efficiency much higher.[5] In addition, this study suggests that by using buried $n^+$-$p^+$ TJs, the Ga-polar structure which is structurally equivalent to N-polar ones can be grown, enabling the use of the attractive properties of N-polar structures, without paying the penalty of the poor optical properties of N-polar (In,Ga)N and sensitivity to wet chemical etching, thus enabling nanostructure fabrication. Buried $n^+$-$p^+$ TJs also make it possible to integrate and stack multiple light emitters such as monolithic multiple color LEDs.[23]

The authors thank Kevin Lee and Henryk Turski for useful discussions. This work was supported by National Science Foundation (NSF) grants: NSF DMREF award #1534303 monitored by Dr. J. Schluter, NSF Award #1710298 monitored by Dr. T. Paskova, NSF CCMR MRSEC Award #1719875, and NSF RAISE TAQs Award #1839196 monitored by Dr. D. Dagenais. The cleanroom fabrication at the Cornell Nanofabrication Facility (CNF) was supported in part by the NSF National Nanotechnology Coordinated Infrastructure (Grant ECCS-1542081).

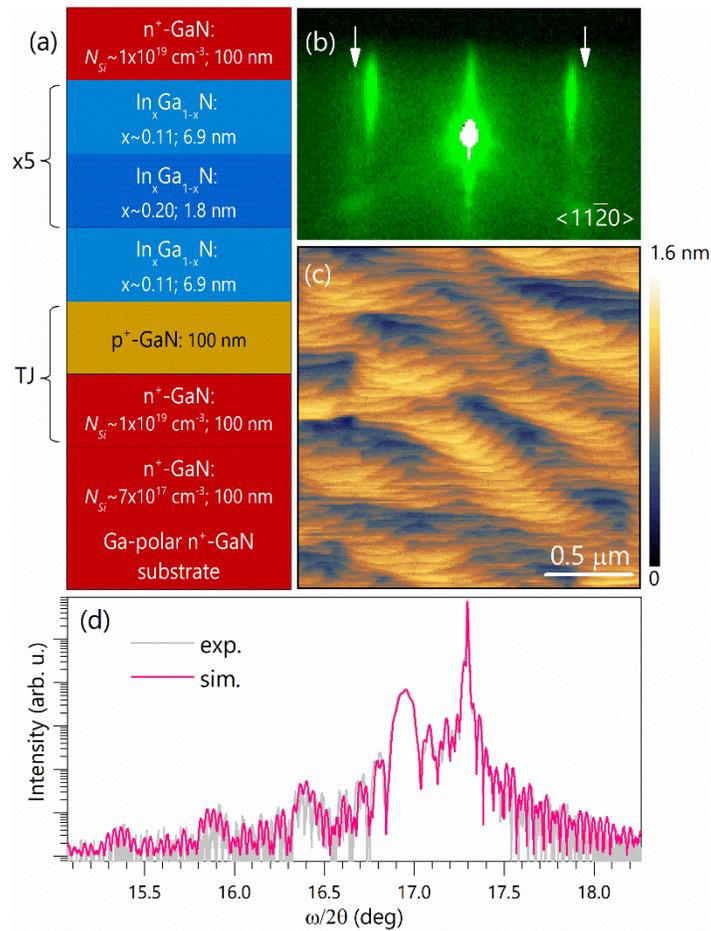

Fig. 1 (a) Schematic layer structure of MBE-grown (In,Ga)N LEDs with a buried $p^+$-GaN layer. (b) The RHEED pattern, (c) 2 × 2 µm$^2$ AFM micrograph and (d) symmetric XRD ω-2θ scan of sample A. The RHEED pattern has been taken below 300 ℃ along the <11-20> azimuth after growth. The two arrows in (b) indicate the RHEED pattern from a Ga-bilayer on the surface. The root-mean-square roughness measured by AFM on the surface in (c) is 0.24 nm.

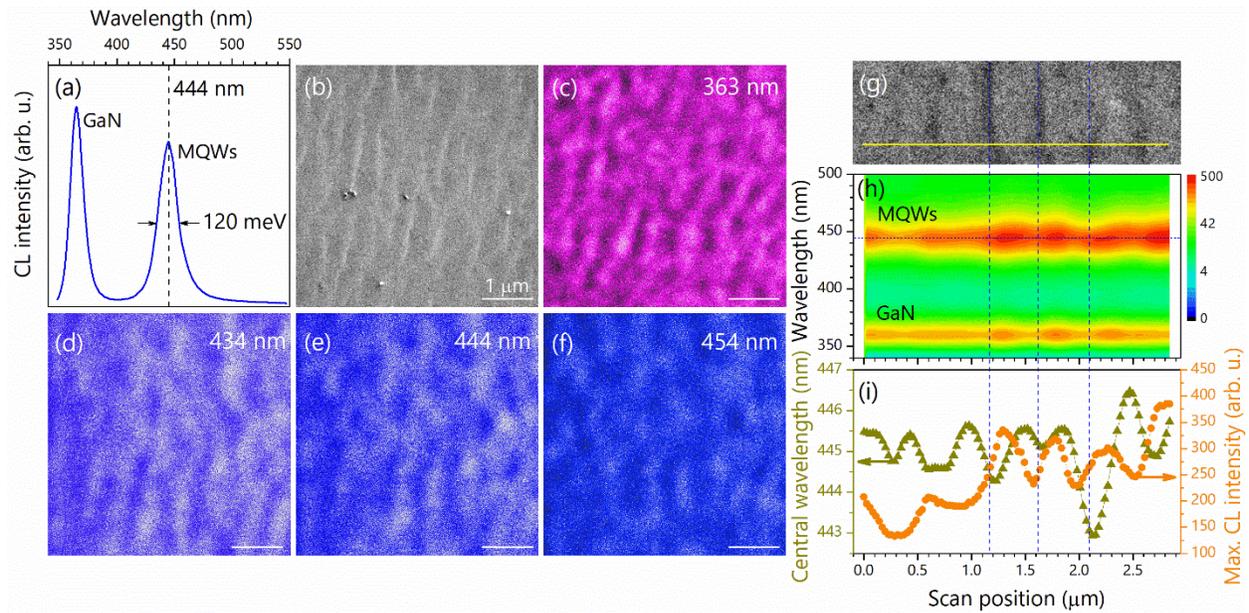

Fig. 2 (a) Cathodoluminescence (CL) spectrum averaged over an area of 88 µm² of sample A. (b) Secondary electron (SE) and (c)–(f) CL images of the same surface region of sample A. For the CL images, the detection wavelength is (c) near-band gap of GaN, (d) on the short-wavelength side, (e) at the center, and (f) on the long-wavelength side of the (In,Ga)N QW peak as indicated by the dashed line in (a). The lengths of the scale bars on (b)–(f) are 1 µm. (g) SE image of a surface piece of sample A. (h) CL line-scan map acquired along the yellow line in (g). Central wavelengths (triangles) and maximum CL intensities (circles) of the (In,Ga)N QW signal in (h) are depicted in (i). The three dashed lines in (g)–(i) indicate the positions of the step terrace edges crossing the scan line in (g).

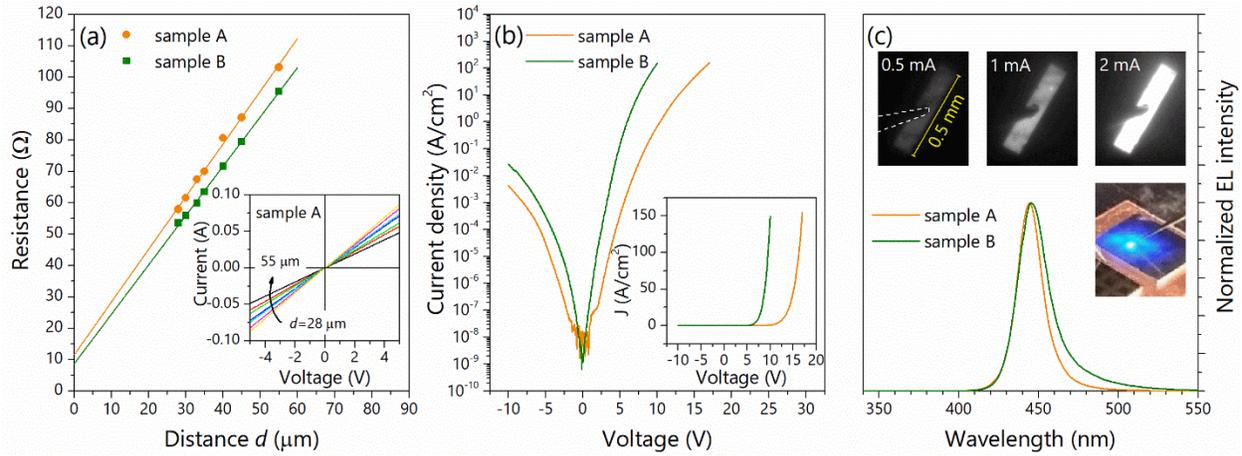

Fig. 3 (a) Resistance as a function of spacing of TLM pads of samples A and B. The lines in (a) are linear fits for extracting top contact resistivities. The inset in (a) is, as a representative, the current vs voltage characteristics of sample A measured on two top $n^+$-GaN contacts separated by distance $d$. Current density vs voltage ($J$-$V$) characteristics of samples A and B in semilog scale (b) and in linear scale [the inset in (b)], where the voltage was applied to the backside of the substrates and the top contacts were grounded. (c) Normalized EL spectra of samples A and B at a constant injection current of 1 mA. The upper three insets in (c) show optical micrographs taken when light is emitted from a device in sample B in a 0.5 × 0.1 mm$^2$ mesa at currents of 0.5, 1 and 2 mA, respectively. The dashed line in the left inset outlines the probe for the top contact, which is much smaller than the mesa, most of the mesa has no metal on top. The lower inset is a color photo at an injection current of 1 mA showing blue emission. The sizes of the devices for the $J$-$V$ in (b) and the EL spectrum in (c) are 80 x 80 μm$^2$.